# Urban Scaling of Cities in the Netherlands

Anthony F.J. van Raan[1*], Gerwin van der Meulen[2], Willem Goedhart[2]

**1** Centre for Science and Technology Studies (CWTS), Leiden University, Leiden, The Netherlands
**2** Decisio | Economic Consulting, Amsterdam, The Netherlands
**\*** corresponding author vanraan@cwts.leidenuniv.nl

## *Abstract*

*We investigated the socioeconomic scaling behavior of all cities with more than 50,000 inhabitants in the Netherlands and found significant superlinear scaling of gross urban product with population size. Of these cities, 22 major cities have urban agglomerations and urban areas defined by the Netherlands Central Bureau of Statistics. For these major cities we investigated the superlinear scaling for three separate modalities: the cities defined as municipalities, their urban agglomerations and their urban areas. We find superlinearity with power-law exponents of around 1.15. But remarkably, both types of agglomerations underperform if we compare for the same size of population an agglomeration with a city as a municipality. In other words, an urban system as one formal municipality performs better as compared to an urban agglomeration with the same population size. This effect is larger for the second type of agglomerations, the urban areas. We think this finding has important implications for urban policy, in particular municipal reorganizations. A residual analysis suggests that cities with a municipal reorganization recently and in the past decades have a higher probability to perform better than cities without municipal restructuring.*

## Introduction and political context

In recent years there is a rapidly growing interest in the role of cities in our global society. Cities are regarded as the main locations of human activity and, particularly, creativity [1]. Population size is an important determinant of the intensity of many socioeconomic [2, 3, 4, 5], infrastructural and knowledge production activities [6, 7, 8] in cities. Indicators representing these activities appear to scale nonlinearly with the number of inhabitants of cities and urban agglomerations. The theoretical basis of this scaling behavior is provided by the theory of complex, adaptive systems [9] in which networked structures reinforce nonlinearly as the system grows, particularly more than proportional, i.e. superlinearly, described by a power law [10]. Also the relation between scaling laws and processes of change in urban systems, for instance the concentration of new technologies and professions is important and could provide understanding of scaling in the context of an evolutionary theory [11]. Moreover, the density of the population – which determines the average distance of interaction- is discussed as an important variable in explaining superlinear scaling [12]. This discussion relates to the problem of the relevant spatial unit of analysis in complex systems. Also for universities superlinear scaling behavior is found in which size is given by the number of publications and the impact of the publications as the dependent variable [13, 14, 15]. Again, distance of interaction appears to play an important role [16, 17].

Recent research in the US on the development of meaningful urban metrics based on a quantitative understanding of cities shows a more than proportional (superlinear) increase of the socio-economic performance of cities with increasing population [2, 3, 4]. A city that is twice as large (in population) as another city can be expected to have a factor of about 2.15 larger socio-economic performance, for instance in terms of gross urban product. This *urban scaling* phenomenon is important for new insights into and policy for urban development and, particularly in the Netherlands, municipal reorganization of urban agglomerations. Different from the usual focus on measures for cutting down expenses, the urban scaling phenomenon opens new vistas toward socio-economic progress. Possible effects could amount to hundreds of millions of euros which means thousands of jobs per year and per urban area [18]. Furthermore, the interpretation of urban scaling laws is important in the discussion on models of urban growth, structure and optimal size of cities and their regions [19, 20].

The US research on urban scaling is about urban areas (MSA's, metropolitan statistical areas) that have grown autonomously to a specific number of inhabitants, regardless of the formal boundaries of municipalities within an urban area. Thus, it is a synchronic, 'static' measurement that has a predicting value for what happens with socioeconomic variables if, for instance, a city (i.e., urban area) doubles in population in the course of time. This is, of course, different from a situation in which a city defined as a municipality and being the central city of the urban agglomeration, doubles in population by a formal reorganization of all municipalities within the urban area into one new municipality.

Nevertheless it is probable that after some time the newly formed city should meet the scaling values as predicted by its new size of population. Crucial is however the interesting policy question: would these scaling values for the doubled population ('created' by municipal reorganization) not already be attained for the urban agglomeration as a whole, simply because the urban agglomeration regardless of the formal municipal boundaries already has this double population?

It is however plausible that in urban agglomerations with a number of different autonomous municipalities --which can in the Netherlands be as high as 8 in urban agglomerations and 15 in the urban areas-- the socio-economic and political cohesion is not optimally. Thus, the reinforcing, non-linear effects of the (central) city dynamics will be hampered. We hypothesize that for these multi-municipality urban areas the scaling rules will apply, but less than what can be expected on the basis of the size of the total population.

Given the above considerations, the main goal of our study is to answer the following related research questions: (1) how do three different modalities of urban systems, namely cities as municipalities and two types of urban agglomerations, scale; (2) does an urban system as one formal municipality perform better as compared to an urban agglomeration with the same population size but with a number of autonomous municipalities within the agglomeration; (3) and if so, can this difference in performance be attributed to less governmental, social, economic and cultural coherence in an agglomeration with a number of autonomous municipalities?

The answer to these above research questions is very relevant for urban policy in all countries where urban systems consist of autonomous municipalities as the results of this work may provide indications for the improvement of the socioeconomic performance of urban systems based on a better governmental coherence. The framework of this analysis is most appropriate for our investigation because the novel element in our work is a clear distinction within major urban systems between three modalities of organization.

The structure of this paper is as follows. First we describe our data and method to investigate the scaling behavior of (1) all cities in the Netherlands (municipalities) with more than 50,000 inhabitants and (2) of the 22 major urban systems in Netherlands (with three modalities: municipalities, urban agglomerations, urban areas). Next we discuss the results and their policy implications. We close by presenting the statistical analyses used in this paper.

In the Supporting Information we provide the results of a preliminary time-dependent analysis of the scaling behavior of cities in the Netherlands as a function of time. This is particularly important in the study of urban growth and scaling of socioeconomic performance.

## Results and Discussion

### Data

We created two sets of cities. First, all cities (defined as municipalities) in the Netherlands with more than 50,000 inhabitants, in total 69 cities; the range of population is 50,000 to the largest city (municipality) with 800,000 inhabitants (Amsterdam). We refer to this set of cities as Set 1. These '50,000+' cities include different types of cities: larger central cities, i.e., major cities that are the centers in a major urban area; smaller central cities in a more countryside-type region; and cities that are suburbs of larger cities.

We collected[1] for all these 50,000+ cities (municipalities) in the Netherlands for the period 2010-2012 the following three variables: (1) number of inhabitants (population); (2) gross urban product in million Euros (index 2013 $\triangleq$ 100); (3) employment (number of jobs). We focus in this paper on the gross urban project (GUP) because the number of jobs correlates strongly with the GUP. Data on the number of jobs are available from the authors. In addition we collected for all cities the land surface areas (in $km^2$, total surface area corrected for water surface area).

Next, within Set 1 (all 50,000+ cities) we focus on specific subset, namely 22 major cities for which the Netherlands Central Bureau of Statistics (CBS)[2] defines two types of agglomerations. First, the *urban agglomeration* which is the central city and the immediately connected suburban cities. Second, the *urban area* in which in addition to the urban agglomeration all other suburban cities that are closely socio-economically connected to the central city are included. The largest urban area, Amsterdam, counts 1.5 million inhabitants.

These 22 major cities are included in Set 1, but we extend this subset with all suburban cities of the 22 major cities. Thus, the novel aspect of this study is that we collected the same data (updated, period 2011-2013) as in Set 1 for (1) the larger central cities themselves (as a municipality, in total 22), (2) their urban agglomerations (in total the 22 central cities and 44 suburban cities), and (3) their urban areas (in total the 22 central cities, the 44 suburban cities in the agglomerations, and additional 90 suburban cities). Moreover, for each of the 134 suburban cities the distances (in km) from their city center to the center of the central city are collected.

We analyzed the scaling of the gross urban product with population for these two sets. In the next sections we discuss the results of the analysis.

### Scaling Analysis: Basic Results

Our main finding is that for both sets of cities the gross urban product scales superlinearly with population. We show the results for Set 1 (all 50,000+ cities) in Fig.1 and the results for Set 2 (the 22 larger cities) in Fig.2. If we use a linear fit of the data, extrapolation would give a negative gross urban product for all cities below 25,000 inhabitants. For a further statistical test see Section 'Materials and Methods'.

---

[1] The source of the data is the Netherlands Central Bureau of Statistics (CBS).
[2] See CBS website http://www.cbs.nl/NR/rdonlyres/6B93A2A7-9A2C-4544-AA9A-7E3E37902778/0/indelingvannederlandinstadsgewesten112014.pdf

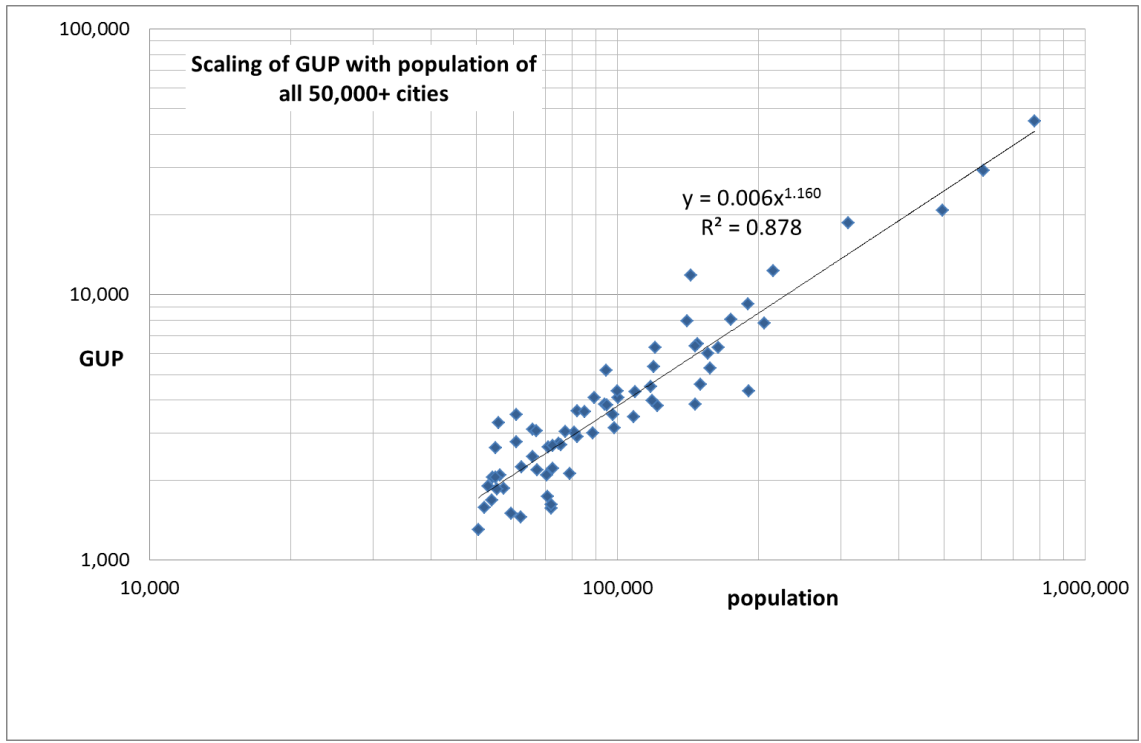

Figure 1. Scaling of the gross urban product (GUP, in million Euros) with population for all Dutch cities above 50,000 inhabitants (Set 1).

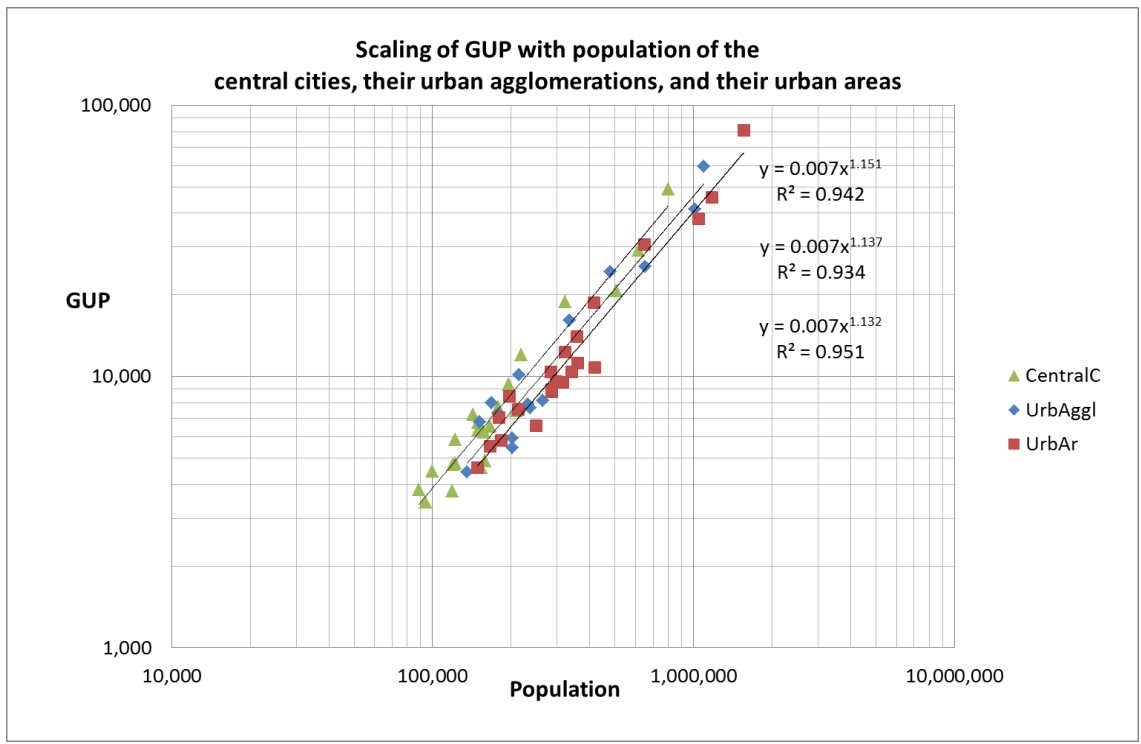

Figure 2. Scaling of the gross urban product (GUP, in million Euros) with population for all the 22 larger central cities (Set 2). Blue diamonds: central cities as municipalities; red squares: urban agglomerations; green triangles: urban areas.

The 50,000+ cities defined as municipalities (Set 1) scale with power-law exponent 1.160 (95%CI: 1.113-1.207) for the gross urban product. The 22 larger central cities (Set 2) scale for the gross urban product with the following exponents: (1) 1.151 for the 22 central cities as a municipality (95%CI: 1.078-1.228); (2) 1.137 for the urban agglomerations of these cities (95%CI: 1.052-1.224); and 1.132 for the urban areas of these cities (95%CI: 1.077-1.207). We discuss the error estimation in Section 'Materials and Methods'. We observe that all three city modalities scale with a power-law exponent around 1.15 with a slight decrease of the exponent from central cities as municipalities, to urban agglomerations and urban areas. Thus our analysis of Set 2, the 22 larger central cities in three modalities gives a positive answer to the first research question about the scaling of these three different modalities.

Moreover and interestingly, the absolute value of the gross urban product for both the urban agglomerations and the urban areas is lower than for the central cities as municipalities. Thus, although both types agglomerations scale with population, they underperform as compared to cities defined as municipalities. We will discuss this important finding further in the Section 'Analysis of the difference in performance between agglomerations and municipalities'. First we continue the analysis of the 50,000+ cities.

## Residual analysis of the 50,000+ cities

We calculated for all cities in the Netherlands above 50,000 inhabitants the residuals of the scaling relation presented in Fig. 1. Residuals are a measure of the deviations of the observed value from the expected value as established by the scaling function. Positive residuals indicate that a city performs better than expected. In Section 'Materials and Method' we discuss the mathematical procedure used to calculate the residuals.

We show in Fig.3 the entire ranking distribution. Then we split the distribution into the ranking of the positive as well as negative residuals. We label the measuring points with the names of the cities concerned, see Figs.4 and 5.

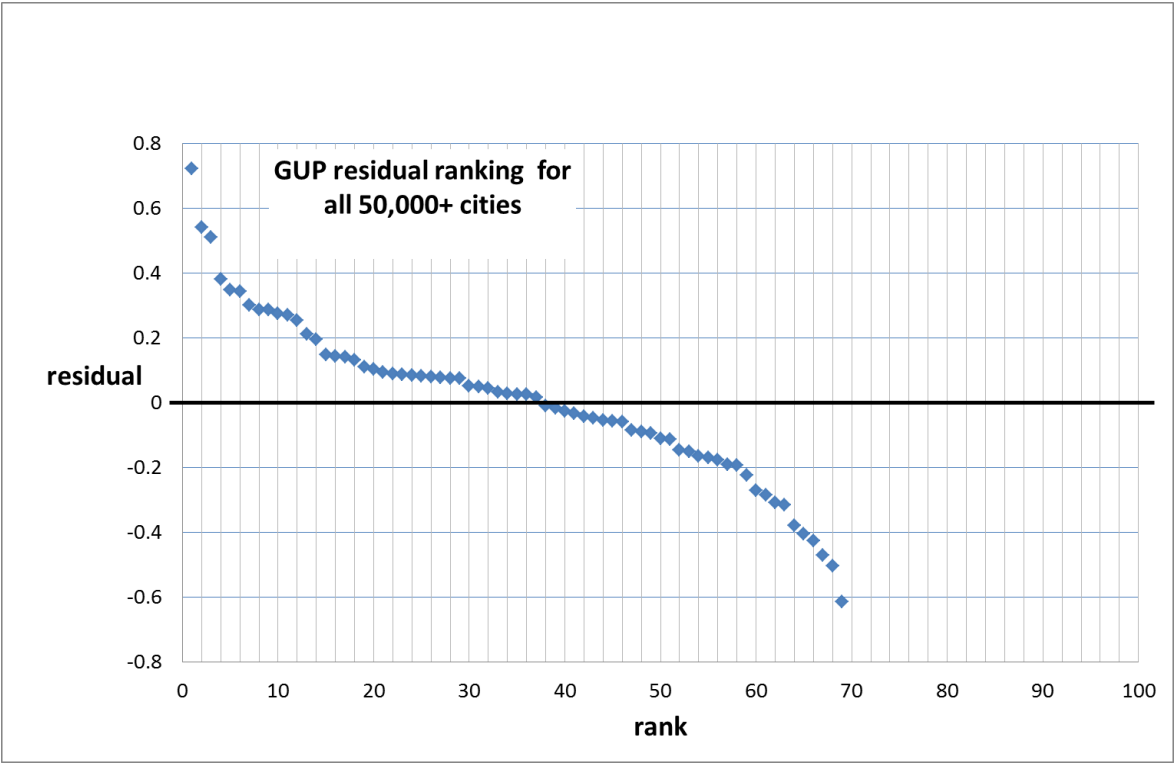

*Figure 3. Ranking of the residuals for the gross urban product (Set 1).*

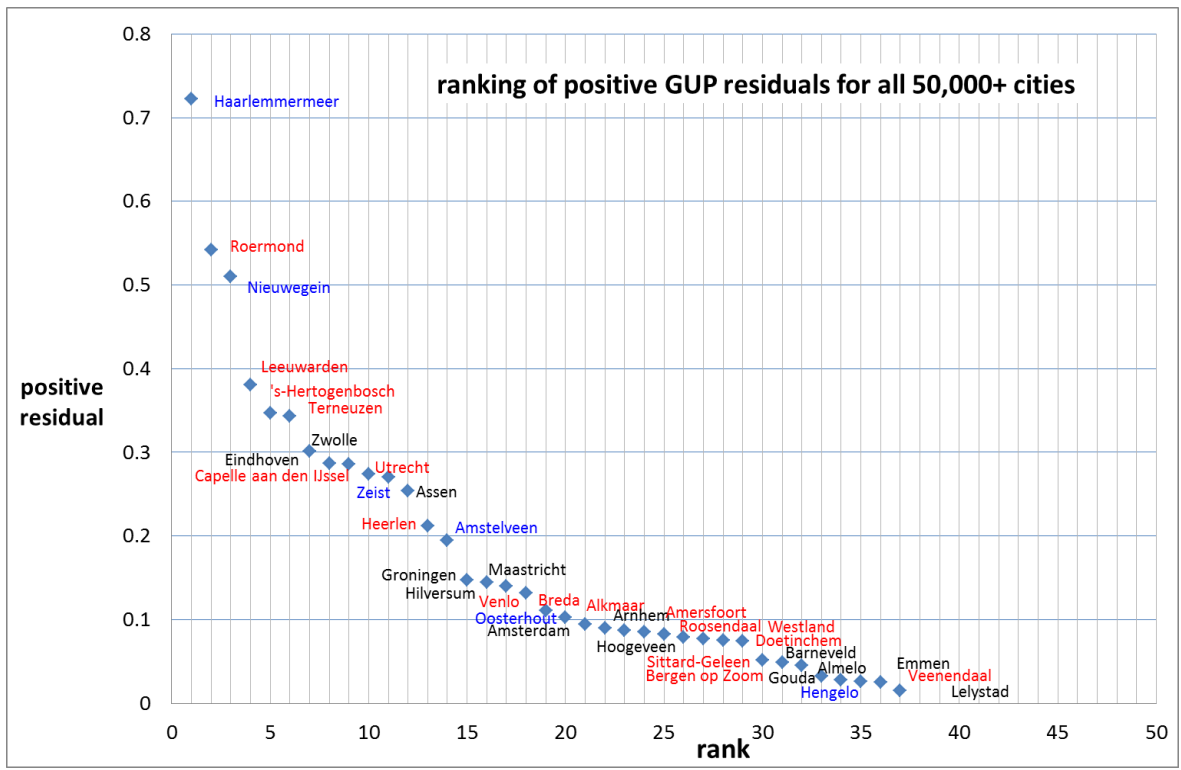

*Figure 4. Ranking of the positive residuals for the gross urban product with the names of the cities. Cities that are suburbs of larger cities are marked blue, cities with a municipal reorganization after 1974 (i.e., fusion of the central city with adjacent smaller cities) are marked red.*

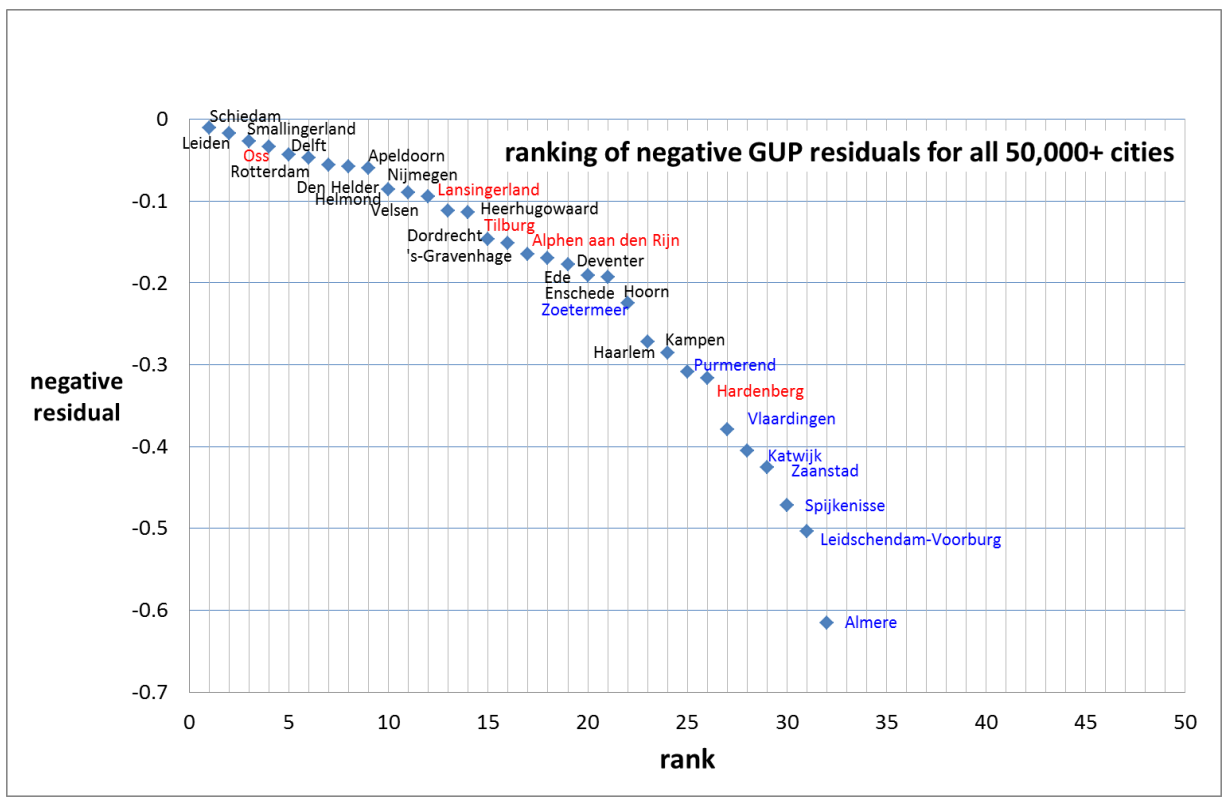

*Figure 5. Ranking of the negative residuals for the gross urban product with the names of the cities. Cities that are suburbs of larger cities are marked blue, cities with a municipal reorganization after 1974 (i.e., fusion of the central city with adjacent smaller cities) are marked red.*

Analysis of the residuals may reveal local characteristics of individual cities in terms of success or failure relative to other cities [3]. As example we mention the two extremes: Haarlemmermeer with the largest positive residual (see Fig.4) and Almere with the largest negative residual (see Fig.5). The relatively bad position of the new city Almere is striking. Clearly this new city, which is part of the Amsterdam urban area does not function (yet) as a 'real' city. Also other suburban cities tend to perform less well. Partly this can be explained by the more residential character of some suburbs, but this is certainly not a general characteristic. Also in the central cities there are large residential areas.

Haarlemmermeer is a remarkable exception. It is also a relatively new city in the Amsterdam urban area. The high positive residual of Haarlemmermeer can be explained very well: Amsterdam International Airport Schiphol, the fourth largest airport in Europe, is located in the municipality of Haarlemmermeer.

We believe that another, and perhaps most important local characteristic is the existence of one coherent governance of central city and its agglomeration instead of several autonomous municipalities within one urban agglomeration. We investigate this in the following way (for more details [18]). From the 69 cities with more than 50,000 inhabitants we removed cities that are suburbs (but autonomous municipalities) within the urban agglomeration or urban area of the major cities in the Netherlands, and also cities that are municipalities in the more rural areas. This means that we focus on the 22 larger central cities with formal urban agglomerations and urban areas as defined by the Netherlands Central Bureau of Statistics[3]. We add three other cities of around 100,000 inhabitants with an informal regional structure[4]. From the analysis of the residuals follows that 15 of these cities have a positive residual and 10 have a negative residual. Of the 15 cities with a positive residual, 9 had in the last decades a municipal reorganization (5 had a substantial reorganization, 2 had a smaller reorganization), and 2 cities with a positive residual had no municipal reorganization. On the other hand, of the 10 major cities with a negative residual no one had a substantial municipal reorganization, one had a smaller reorganization and 4 cities with a negative residual had no municipal reorganization.

This analysis of Set 1 suggest that cities with a municipal restructuring recently and in the past decades have a higher probability to perform better than cities without municipal restructuring [18]. Thus we find along this line at least a first indicative answer to our second research question whether an urban system as one formal municipality performs better as compared to an urban agglomeration with the same population size but with a number of autonomous municipalities within the agglomeration.

## Analysis of the difference in performance between agglomerations and municipalities

As discussed in the foregoing section, our analysis of Set 2, the 22 larger central cities in three modalities which are central cities as municipalities, urban agglomerations, and urban areas, gives a positive answer to the first research question about the scaling of these three different modalities. We showed in the foregoing section that the gross urban product for all three modalities scale with a power-law exponent of around 1.15 with possibly a slight decrease of the exponent from central cities as municipalities, to urban agglomerations and urban areas. But we also observe (Fig.2) that the absolute value of the gross urban product for both the urban agglomerations and the urban areas is lower than for the central cities as municipalities[5]. Thus, although both types agglomerations scale with population, they underperform as compared to cities defined as municipalities. This difference in performance is significant, as shown in Section 'Materials and Methods'. The ratio of central city and urban agglomeration performance for the same population size (measured by the expected GUP) is 0.86, and the ratio of central city and urban area performance is 0.76. This observation answers in the affirmative the second research question whether an urban system as one formal municipality performs better as compared to an urban agglomeration with the same population size but with a number of autonomous municipalities within the agglomeration.

---

[3] These cities (with population of the urban areas rounded to the nearest 1000) are: Amersfoort (286,000), Amsterdam (1,529,000), Apeldoorn (213,000), Arnhem (358,000), Breda (318,000), Dordrecht (264,000), Eindhoven (412,000), Enschede (315,000), Den Haag (The Hague) (1,028,000), Groningen (352,000), Haarlem (414,000), Heerlen (251,000), 's Hertogenbosch (195,000), Leeuwarden (162,000), Leiden (338,000), Maastricht (181,000), Nijmegen (283,000), Rotterdam (1,170,000), Sittard-Geleen (150,000), Tilburg (295,000), Utrecht (626,000), Zwolle (178,000).
[4] These cities are: Alkmaar, Deventer, Venlo.

The third research question concerns the attribution of the above discussed difference in performance to less governmental, social, economic and cultural coherence in an agglomeration with a number of autonomous municipalities. Before answering this last and politically important research question, a prior question arises: is it to be expected that agglomerations will underperform as compared to the central cities because in agglomerations the urban structure will probably be less dense than the central cities and thus the benefits of scaling may be lower? We find however that the densities of the central cities as municipalities are spread out over a wide range (average for the 22 central cities is 2506 ± 1574 (*sd*) inhabitants/km$^2$) and that the densities of the urban agglomerations (average for the 66 cities in the urban agglomerations is 2047 ± 741 (*sd*) inhabitants/km$^2$) lie completely within the density range of the central cities and thus their differences are statistically not significant. This is understandable: in several cases a central city and its urban agglomeration are the same because the agglomeration has been merged into the central city as one municipality, whereas in other cases with a similar agglomeration structure the suburbs are still separate municipalities. The densities of the urban areas are lower (average for the 156 cities with the urban areas is 1159 ± 588 (*sd*) inhabitants/km$^2$) but still partly within the density range of the central cities. Thus, it is unlikely that differences in population density, particular between central cities and their urban agglomerations can explain the above discussed underperformance.

There is a further argument against the use of population densities to explain the differences in performance between central cities and their agglomerations. Densities per city are averages calculated on the basis of the entire land surface area of cities. In many cases, suburbs immediately connected to the central city have a large surface area (in quite a number of cases larger than the surface area of the central city) located in a direction away from the city. The use of the average population density of such suburbs suggests a spread of the population over the entire surface area. As discussed, this is often not the case: the large majority of population is concentrated in the suburban city bordering directly on the central city, and the rest of the surface area belonging to the suburban city is much more sparsely populated as it consists of woods, polder land etc.

We also performed another test to find out whether densities could play a role in the performance of cities. We calculated for the 22 central cities the residuals of the scaling relation between the gross urban product and populations size. We find a very weak and, remarkably, a negative correlation (r= -0.191) between residuals and population densities.

The above shows that the use of population densities to explain the relative underperformance of urban agglomerations and urban areas is not appropriate. It is attractive to consider urban agglomerations and urban areas as a network structure of connected cities and to explain their underperformance in terms of (de)centrality. There are many definitions of centrality [21] in order to optimize this measure depending on the type of network. Generally, centrality is a measure to identify and rank the most important nodes in a network and to assess the role of nodes in the cohesiveness of the network. The larger the network structure, the higher the probability that nodes other than the original central node will decrease the importance of the original central node. In the case of an urban network this would imply that an urban structure such as an agglomeration is less cohesive than a compact central city. Thus the agglomeration will underperform as compared to a compact central city with the same population because socioeconomic performance largely depends on cohesiveness. However, the network models used for the calculation of centrality consist of nodes of 'equal size', for instance, all the nodes are individuals, publications, citations. But this is evidently not the case in urban agglomerations.

---

[5] The gross urban product for an urban agglomeration or urban area is the sum of the gross urban products of the central city and the suburban cities in the urban agglomeration or urban area.

Therefore, instead of applying network centrality measures we used a model for the closeness of central city and suburban city which is based on the center of mass principle in physics. The center of mass is the mean location of a distribution of mass in space. In our approach, mass is the population of the cities within an agglomeration. We calculate for each suburban city separately its center of mass location with respect to the central city, or to the urban agglomeration as a whole in the case of suburban cities in the urban area. Because we know the surface area of the central city we can calculate the diameter $D$ and hence the radius $D/2$ of the central city size by assuming that the shape of the city is approximately circular. Of all 134 suburban cities we collected the distance $r_i$ of the suburban city $i$ to the center of its central city ($i=1$) normalized to the radius $D/2$ of the central city surface area. Of these 134 suburbs 44 belong to the urban agglomerations (first 'ring' of suburbs around the central cities). The urban areas consist of the urban agglomerations with in addition 90 suburbs in a second 'ring' around the central city. On average, an urban agglomeration consists of a central city and 2 suburban cities, and an urban area (which always includes the urban agglomeration) consists of a central city and 6 suburban cities (including the 2 within the agglomeration).

We analyze the structure of the urban agglomerations as follows. By putting the center of the central city in the origin of the coordinate system, we calculate the location of the center of mass $R_i$ for a suburb $i$ in the urban agglomeration:

$$R_i = \frac{N_i}{N_1} r_i$$

(1)

where $N_i$ is the population of the suburban city and $N_1$ is the population of the central city.

If $R_i$ is 1 the center of mass lies at a distance equal to the central city radius $D/2$ from the central city center, i.e., on the border of the central city. For any value of $R_i < 1$ the center of mass lies within the central city. Theoretically, a small city could be far away (say, 100 km) from a large city whereas the center of mass could still lie within the larger city. No one would consider this situation as an urban agglomeration. Therefore we apply a second criterion by requiring that the distance $r_i$ of the suburban city to the center of the central city must be smaller than at most $D/2$ from the border of the central city, i.e., $r_i \leq 2$. We find for the urban agglomerations an average $<R_i> = 0.20$ ($sd=0.17$) and an average $<r_i> = 1.24$ ($sd=0.45$). All 44 suburban cities in the urban agglomerations satisfy the center of mass criterion and 91% satisfy both criteria. The remaining 9% has $r_i$ values between 2.10 and 2.56. This means that in the urban agglomerations by far the most suburban cities are directly bordering to the central cities and form one compact urban system with the central city.

Given the above findings we consider the central city and its agglomeration suburbs (the first 'ring' of suburbs) as one compact city and calculate the center of mass for the 90 additional suburban cities in the urban areas (the second 'ring' of suburbs):

$$R_i = \frac{N_i}{\sum_{i=1}^{n} N_i} r_i$$

(2)

where $N = \sum_{i=1}^{n} N_i$ is the total population of the urban agglomeration. For the urban areas we require that the distance of the suburban city to the center of their central city must be smaller than at most the central city diameter $D$ from the border if the central city, i.e., $r_i \leq 3$. Excluding the 44 suburban cities that belong to the urban agglomeration (the first 'ring' of suburbs around the central city) and thus focusing on the 90 additional suburban cities (the second 'ring') we find an average $<R_i> = 0.25$ ($sd=0.19$) and an average $<r_i> = 2.01$ ($sd=0.78$). All these 90 suburban cities in the urban areas satisfy the center of mass criterion and 86% satisfy both criteria. The remaining 14% has $r_i$ values between 3.03 and 4.22. This means that in the urban areas a large majority of the suburban cities is directly bordering to the first ring of suburbs and contributes to the compact urban system around the central city. These findings can easily be verified by observing the urban agglomerations with Google Maps.

With the above discussion we have shown that it is unlikely that density and centrality arguments can entirely explain the differences in performance between central cities as municipalities and the urban agglomerations and urban areas of these cities. These findings give an answer to our third research question: we believe that these differences are to at least a considerable extent the consequences of a less governmental, social, economic and cultural coherence in an agglomeration with a number of autonomous municipalities.

## Conclusions and policy implications

In most earlier work on urban scaling the 'cities' are in fact larger agglomerations around central cities. It is emphasized [3] that these agglomerations are socioeconomic units and therefore the defining feature of cities, this in contrast to administrative definitions which are regarded as more arbitrary.

We however consider that the governmental definition of cities within the direct urban region around a central city does matter because these definitions often have very longstanding and deep historical, political and social grounds. We think that this is also the case in, for instance, the US, but it is certainly the case in the Netherlands. The novel and unique property of our study is precisely that we are able to distinguish three well-defined urban modalities. Certainly, a city definition on the basis of socioeconomic considerations is important and we emphasize that the definitions used in this study for the urban agglomerations and the urban areas are indeed based by the Netherlands Bureau of Statistics on socioeconomic connections between the central cities and their agglomerations.

But although urban agglomerations are characterized by socioeconomic connections, this does not mean that the governance structure within these agglomerations has a strong cohesiveness resulting in an optimal social, economic and cultural coherence. Quite the contrary, the urban agglomerations and urban areas consist of independent, autonomous municipalities each having their own political and social agenda. For instance, even a medium-sized compact urban area such as Leiden consists of ten autonomous municipalities for about 350,000 inhabitants. Every four years there are in the Netherlands new municipal elections which may involve a complete change of political orientation. This often results in new policy making in which previous partnership within the agglomeration may be revised or even eliminated thereby eroding the culture of mutual confidence. As a consequence, the urban agglomeration may suffer from the lack of vigour and perseverance in the realization of infrastructural, cultural and economic (particularly industrial business areas) facilities.

In summary, we have investigated the scaling behavior of cities and their agglomerations. The framework developed in this study leads to challenging conclusions about the importance of a one-municipality instead of a multi-municipality governance in major urban agglomerations. A coherent governance of major cities and their agglomerations may create more effective social interactions which reinforce economic and cultural activities generating a substantial wealth benefit. Even if not all of the differences in performance between central cities and their urban agglomerations and urban areas can be explained by incoherent governance, then still a substantial part of the above indicated benefits would generate a significant increase of wealth and disposable resources. If the benefit would be only 10% of the expected value, then still we are talking in terms of 100 million Euros per city resulting in thousands of jobs.

We believe that an important next step in this study of urban scaling and the consequences for urban governance is a comparison of these results for Netherlands with other countries in the European Union. Examples are Denmark where in 2007 a drastic municipal reorganization of all cities and their agglomerations was implemented, and to a substantial extent also Belgium and Germany.

## Materials and Methods

### Definitions of municipalities, urban agglomerations, and urban areas

The source of the data and of the definitions of the urban agglomerations and urban areas is the Netherlands Central Bureau of Statistics (CBS). The data on the land surface areas are taken from the Wikipedia websites of the cities and the distances of the suburban cities to the central cities are determined with an automated version of a distance table[6].

### Calculation of the residuals and statistical tests

We calculated the residuals of the power-law scaling of the gross urban product with population for the analysis of the real performance as compared to the expected value, see Section 'Residual analysis of the 50,000+ cities'. The mathematical procedure is presented in the text box here below. The residuals are also used to test the heteroskedasticity (see for instance [22]).

> A power-law relation between for instance the gross urban product ($G$) and population ($P$) can be written as:
>
> $$G(P) = aP^\beta \qquad (3)$$
>
> We find empirically (as an example see Fig. 1, Set 1) the value 0.006 for the coefficient $a$ and 1.160 for the power-law exponent ß.
>
> Denoting the *observed* value of the gross urban product for each specific city with $G_i$ we calculate the residuals $\xi_i$ of the scaling distribution of each city as follows [3, 14]:
>
> $$\xi_i = ln[G_i/G(P)] = ln[G_i/aP^\beta] \qquad (4)$$

---

[6] See for instance www.nl.afstand.org.

In Fig.6 we plotted the residuals of the gross urban product against the population for each city with more than 50,000 inhabitants (Set 1). There appears to be no correlation between the magnitude of the residuals and the value of the independent variable, in this case the population. In other words, the figure does not suggest the presence of heteroskedasticity and we can assume that the power-law fit and the inferences drawn from it are valid.

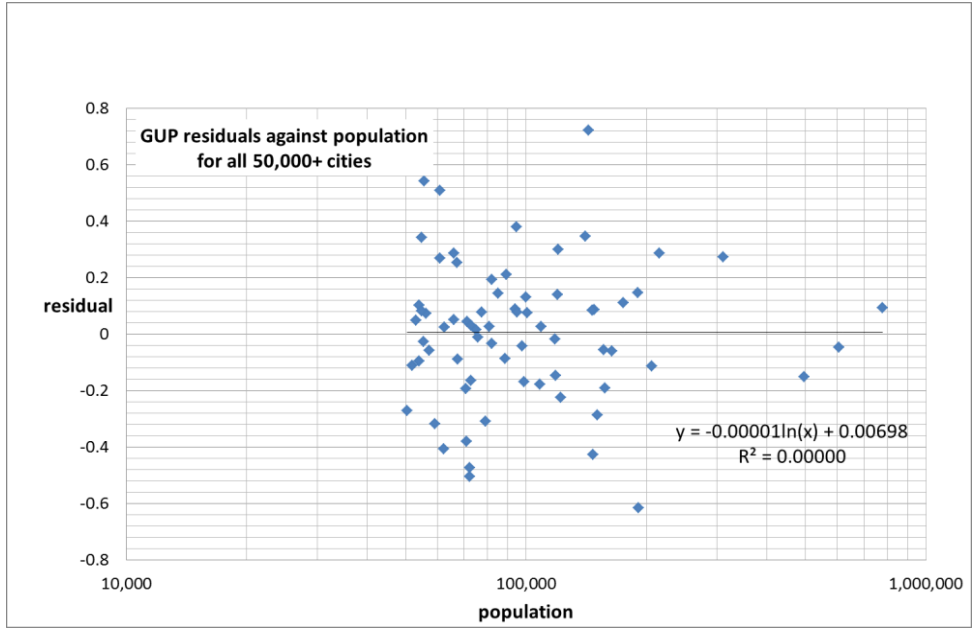

*Figure 6: Plot of the gross urban product residuals against population for all 50,000+ cities.*

We find that the variation in the residuals follows an exponential distribution mirrored by the regression line and can be described by a Laplace exponential distribution density function, defined as [3]:

$$\Phi(\xi) = \frac{1}{2s} e^{(-\frac{|\xi|}{s})} \qquad (5)$$

with parameter *s* which characterizes the width of the distribution and is defined by the mean expectation for the absolute value of the residuals

$$s = <|\xi|> \qquad (6)$$

Fig.7 shows the Laplace distribution function compared with the normalized frequency distribution of the residuals.

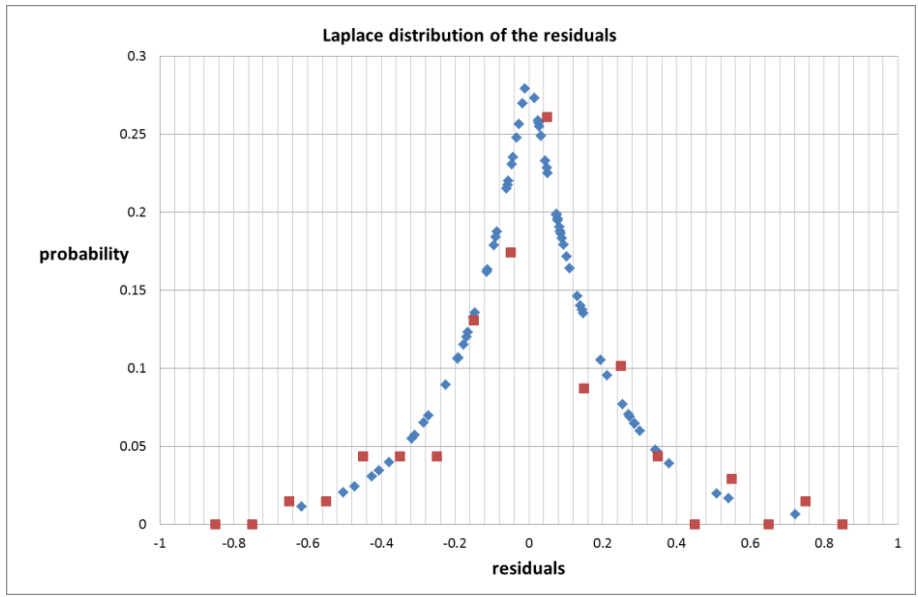

*Figure 7: Laplace distribution of the residuals for the gross urban product (Set 1) compared with their frequency distribution (normalized to the same scale).*

## Error intervals

The relation between GUP and population size is characterized by superlinearity described with a power-law dependence, an indication of cumulative advantage. This is different from a simple regression of, for instance, retail sales and disposable income of households. Therefore we approached the error estimation for the power-law exponents with the following empirical procedure.

In Set 1 (the 69 cities with more than 50,000 inhabitants) we removed 10 times randomly 20% of the cities. After each removal of this 20% we calculated the power-law exponent. Next we calculated the average value of these exponents and the standard deviation. The complete set of 69 cities gives an exponents of 1.160, as shown in Fig.1. The average value of the 10 randomized measurements is 1.158 with sd=0.023 and a 95% confidence interval (CI) 1.113-1.207.

For the 22 major cities (Set 2) we followed the same procedure. In all three modalities, the central cities, their urban agglomeration and their urban areas we remove 10 times randomly 20% of the cities, agglomerations and areas. The complete set of 22 cities has an exponent of 1.151, as shown in Fig.2 (curve with triangles). The average value of the 10 randomized measurements is 1.153 with sd=0.036 and the 95% CI is 1.078-1.228. For the urban agglomerations the exponent is 1.137 (Fig.2, curve with diamonds) and the average value of the 10 randomized measurements is 1.138 with sd=0.040, the 95% CI is 1.052-1.224. For the urban areas the exponent is 1.132 (Fig.2, curve with squares) and the average value of the 10 randomized measurements is 1.133 with sd=0.027, the 95% CI is 1.077-1.189.

In order to find out whether the differences in expected gross urban products between the central cities, their urban agglomerations and their urban areas are significant (the mutual distances of the curves in Fig.2), we calculated the confidence intervals for these scaling curves. As an example we performed the calculation for population size N = 200,000. The differences are indeed significant as is shown in Fig.9.

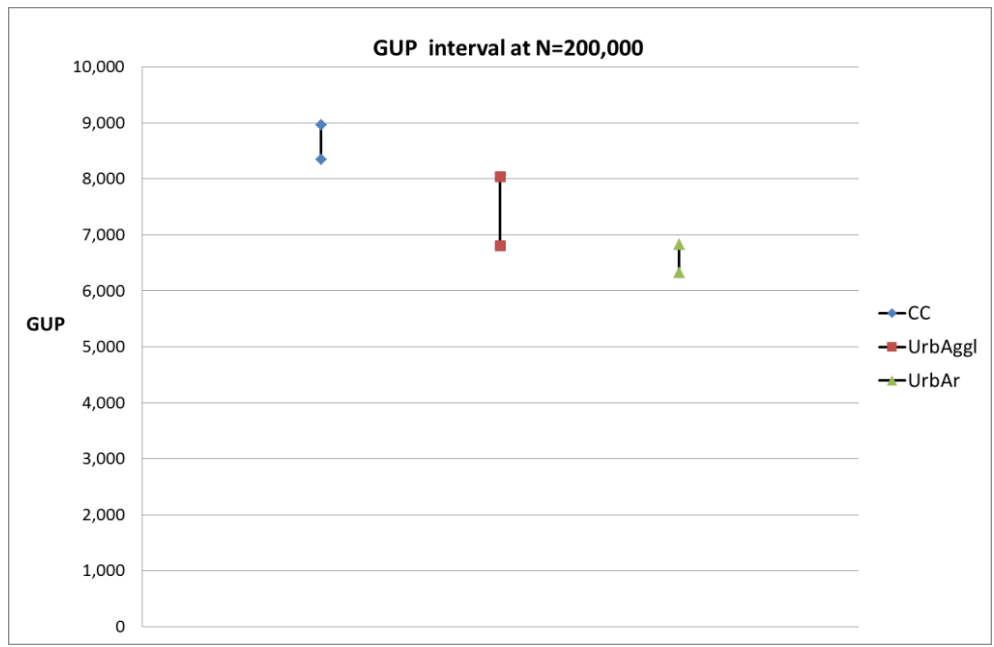

*Figure 8. Confidence intervals for the expected gross urban product (GUP, in million Euros) for the central cities, their urban agglomerations and their urban areas.*

# References

1. Florida R (2004). *Cities and the creative class*. New York: Routledge.

2. Bettencourt LMA, Lobo J, Helbing D, Kühnert C, West GB (2007). Growth, innovation, scaling, and the pace of life in cities. *Proc Natl Acad Sci USA* 104, 17: 7301-7306.

3. Bettencourt LMA, Lobo J, Strumsky D, West GB (2010). Urban Scaling and Its Deviations: Revealing the Structure of Wealth, Innovation and Crime across Cities. *PLoS ONE* 5, 11, e13541.

4. Lobo J, Bettencourt LMA, Strumsky D, West GB (2013). Urban scaling and the production function for cities. *PLoS ONE* 8, 3, e58407.

5. Schläpfer M, Bettencourt LMA, Grauwin S, Raschke M, Claxton R, Smoreda Z, West GB, Ratti C (2014). The scaling of human interactions with city size. *Journal of the Royal Society Interface* 11, 98, 20130789. Preprint available online at http://arxiv.org/abs/1210.5215.

6. Arbesman S, Kleinberg JM, Strogatz SH (2009). Superlinear scaling for innovation in cities. *Phys. Rev* E 68, 066102.

7. Bettencourt LMA, Lobo J, Strumsky D (2007). Invention in the city: Increasing returns to patenting as a scaling function of metropolitan size. *Research Policy* 36, 1007-120.

8. Nomaler O, Frenken K, Heimeriks, G (2014). On scaling of scientific knowledge production in U.S. metropolitan areas. *PLoS ONE* 9, 10, e110805.

9. Holland JH (1995). *Hidden Orders. How Adaptation Builds Complexity*. New York: Basic Books.

10. Bettencourt LMA (2013). The Origins of Scaling in Cities. *Science 340*, 1438-1441.

11. Pumain D, Paulus F, Vacchiani-Marcuzzo C, Lobo J (2006). An evolutionary theory for interpreting urban scaling laws. *Cybergeo* 343, 1-21. Available online at https://cybergeo.revues.org/2519.

12. Pan W, Ghoshal G, Krumme C, Cebrian M, Pentland A (2013). Urban characteristics attributable to density-driven tie formation. *Nature Communications* 4, June 4, 2013, article nr 1961; doi:10.1038/ncomms2961.

13. van Raan AFJ (2006). Performance-related differences of bibliometric statistical properties of research groups: cumulative advantages and hierarchically layered networks. *Journal of the American Society for Information Science and Technology* 57 (14), 1919-1935.

14. Van Raan AFJ (2013). Universities Scale Like Cities. *PLoS ONE* 8, 3, e59384.

15. Jamtveit B, Jettestuen E, Mathiesen J (2009). Scaling properties of European research units. *Proceedings of the National Academy of the United States of America (PNAS)* 106, 32, 13160-13163.

16. van Noorden R (2010). Love thy lab neighbour. *Nature* 468, 1011

17. Lee K, Brownstein JS, Mills RG, Kohane IS (2010). Does Collocation Inform the Impact of Collaboration? *PLoS ONE* 5, 12, e14279.

18. Van der Meulen G, Goedhart W, van Raan AFJ (2014). *Urban scaling van Nederlandse steden*. Report 18-12-2014 for the Netherlands Ministry of the Interior and Kingdom Relations, The Hague (in Dutch, available online at http://www.rijksoverheid.nl/documenten-en-publicaties/rapporten/2014/12/18/urban-scaling-van-nederlandse-steden.html).

19. Putain D (2004). *Scaling laws and urban systems*. SFI Working Paper 2004-02-002, Santa Fe institute. Available online at: http://www.santafe.edu/media/workingpapers/04-02-002.pdf.

20. Putain D (2012). Urban Systems Dynamics, Urban Growth and Scaling Laws: The Question of Ergodicity. In Portugali J, Meyer H, Stolk E, Tan E (eds.): *Complexity theories of cities have come of age: an overview with implications to urban planning and design.* Heidelberg, Berlin: Springer.

21. Martin T, Zhang X, Newman MEJ (2014). Localization and centrality in networks. *Phys. Rev. E*, **90**, 052808. Preprint available online at http://arxiv.org/abs/1401.5093v2.

22. Newbold P (1995). *Statistics for business and economics*. Englewood Cliffs (NJ): Prentice-Hall, fourth edition, p.566-572.

## Supporting Information

Figure S1:

Slowly growing cities: as an example the scaling of the gross urban product (GUP) of Eindhoven (city/municipality, blue diamonds; urban area, red squares) with the number of inhabitants.

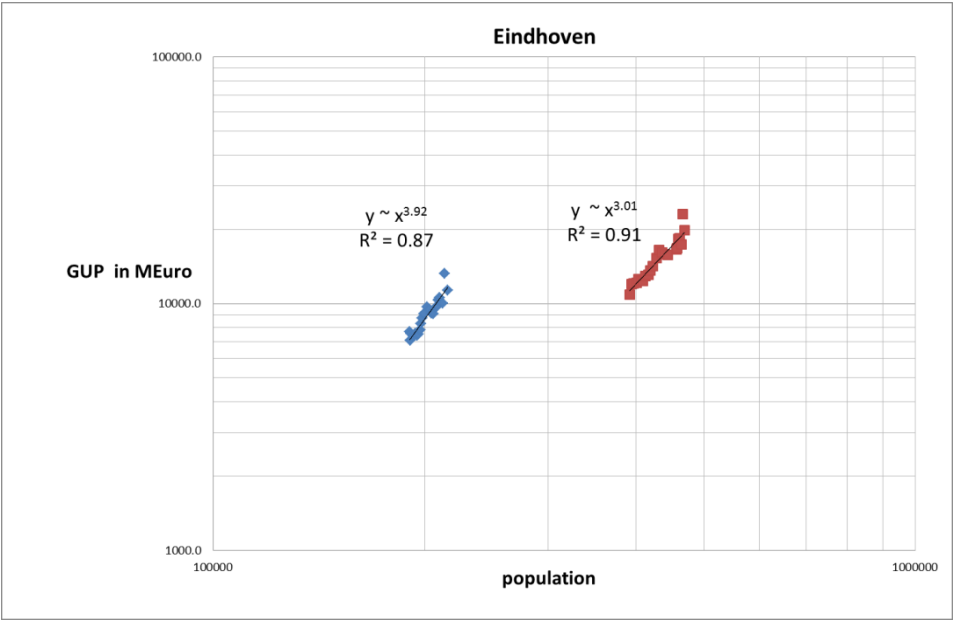

Figure S2:

Rapidly growing cities: as an example the scaling of the gross urban product (GUP) of Almere (city/municipality, blue diamonds) with the number of inhabitants.

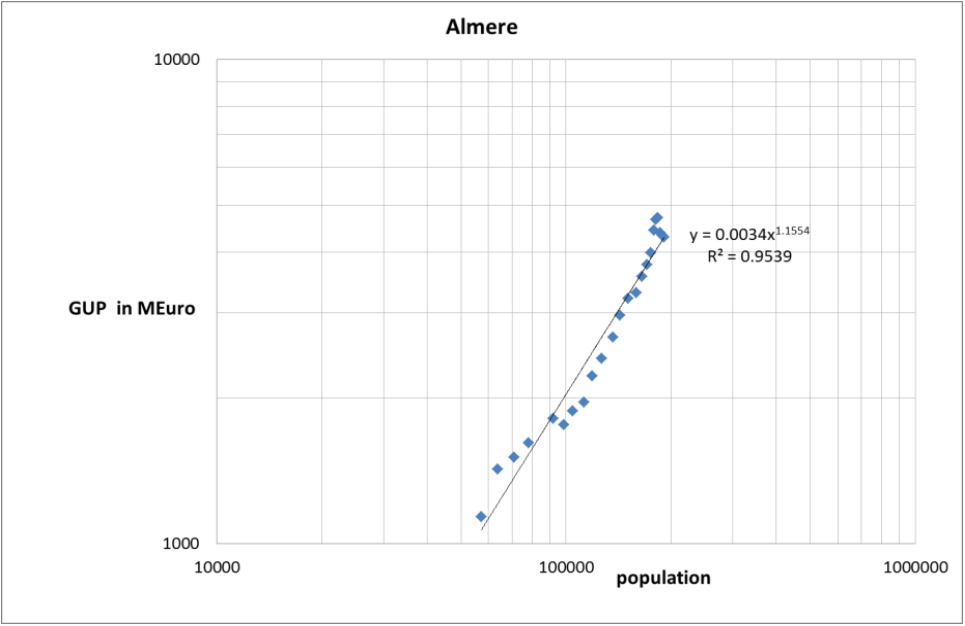

Text S1:

Time-dependent analysis

We collected for 17 cities (defined as municipalities) as well as for 8 of them the agglomerations (which as discussed above mostly consists of several autonomous municipalities) covering in total 62 cities for the period 1988-2013 (1) the number of inhabitants (population); (2) the gross urban product (index 2013 ≙ 100); and (3) employment (number of jobs). As in the main text we focus on the gross urban project (GUP) because the number of jobs correlates strongly with the GUP.

We analyzed the gross urban product as a function of population (if applicable: before and after municipal restructuring, otherwise the city as well as its agglomeration). In this time-dependent ('diachronic') analysis cities are compared with themselves over a period of about 25 years. Because we investigate the scaling behavior of cities on the basis of their population, a time-dependent analysis of a specific city will only yield reliable results if there is a considerable population growth in the time period considered. Thus, there is a major difference between cities that have grown rapidly in the last 25 years, and cities that have increased only moderately or even decreased in population. We first focus on the latter case. Here a statistically significant measure of the gross urban product as a function of population is hardly or not possible. For instance, the acquisition of a major company or the opposite, business closure, may change significantly the number of jobs in a city while the population of the city does not change. Thus, scaling of the gross urban product as a function of population is not applicable.

Generally one can expect that, because of the continuous reinforcing of wealth in a country, the gross urban product of cities will increase slowly as a function of time with an average annual rate (in the Netherlands) of 4.3 per cent in the period 1987-2012. In the case of cities with a slow increase of population, this situation will be characterized by an 'artificially' very large power-law exponent for the gross urban product as a function of population, as explained mathematically below (see textbox at the end of this section). As an example we show in Fig.S1 the scaling of the gross urban product and population of Eindhoven, one of the major industrial centers (Philips, ASML) in the Netherlands. In the period covered by our study (1988-2013) the population of the city (municipality) of Eindhoven (around 200,000 inhabitants) increased with 14 percent and the population of the Eindhoven urban area (around 450,000 inhabitants) with 21 per cent. The GUP (inflation corrected, index 2013) however increased in the same period for the city with 60 percent and for the urban area with 82 percent. As a consequence, we find very high exponents: 3.92 for the city and 3.01 for urban area. Thus it is clear that for cities with a small increase in population the measurement of the scaling behavior of socioeconomic variable with population and resulting exponents is meaningless.

However, cities that have grown rapidly in the covered time period the scaling has lower exponents. We present as an example Almere, the new city in the Amsterdam urban area that more than tripled in population from 60,000 in 1988 to about 200,000 now. We find a scaling power-law exponent of 1.16 for the gross urban product, see Fig.S2. In this case of a rapidly growing city, it is interesting that we find an exponent of 1.43 for the number of jobs. Time analysis of the ratio of GUP and number of jobs shows that the number of jobs increased more rapidly than the GUP. This means a decreasing added value for the more recent jobs. In Section 'Residual Analysis of the 50,000+ cities' further evidence is found that this new city underperforms considerably as compared to other cities, in agreement with earlier socio-economic studies of cities in the Netherlands.

Other fast growing cities show similar and even higher power-law exponents. These fast growing cities are characterized by a typical residential role in the beginning (most of these fast growing cities are within the larger urban areas of major cities). But as these cities became larger they started to attract business companies and thus reinforced their socioeconomic position in a relatively great pace, often in a situation where the population was not growing that rapidly anymore. So also in these cases, the high superlinear exponents can be explained, at least a part of it, by the similar mathematical model as discussed in the text box.

We conclude that the diachronic analysis of cities indeed reveals scaling behavior, but only in the case of a substantial increase of population, for instance rapidly growing new cities. For slowly growing cities with a relative strong increase of GUP we find very high exponents, as illustrated by the Eindhoven case. It illustrates the difficulty to investigate the scaling behavior with time series data.

---

*Mathematical explanation of large power-law exponents*

Suppose we have a slowly growing gross urban product

$$G(t) \sim e^{\alpha t} \tag{S1}$$

and an even slower growing population

$$P(t) \sim e^{\beta t}, \ \beta < \alpha \tag{S2}$$

Say we are interested in the gross urban product as a function of P, then from the necessary condition

$$\int G(t)dt = \int G(P)dP \tag{S3}$$

and using Eq. S2 from which follows $t \sim \frac{1}{\beta} \ln P$ we find

$$G(P) = G(t)\frac{dt}{dP} \sim \frac{e^{\alpha t}}{\beta e^{\beta t}} = \frac{1}{\beta} e^{(\alpha - \beta)\left(\frac{1}{\beta}\right)\ln P} = \frac{1}{\beta} P^{\left(\frac{\alpha}{\beta} - 1\right)} \tag{S4}$$

which is a power-law with exponent γ = (α/β − 1).

It is now obvious from Eq. S4 that in the case of a very slow increase in population (small β, typically 0.003 in our study) and a general increase of the gross urban product with a larger exponent (α, typically 0.025 in our study), the power-law exponent of the gross urban product as a function of population will be considerably higher than 1.

---